\documentclass[manuscript]{aastex}

\begin{document}

\title{On the Binding Energy Parameter $\lambda$ of Common Envelope Evolution}

\author{Xiao-Jie Xu$^{1,2}$ and Xiang-Dong Li$^{1,2}$}
\affil{$^1$Department of Astronomy, Nanjing
University, Nanjing 210093, China;\\
$^2$Key laboratory of Modern Astronomy and Astrophysics (Nanjing
University), Ministry of Education, Nanjing 210093,
China\\lixd@nju.edu.cn}

\begin{abstract}
The binding energy parameter $\lambda$ plays an important role in
common envelope (CE) evolution. Previous works have already pointed
out that $\lambda$ varies throughout the stellar evolution, though
it has been adopted as a constant in most of the population
synthesis calculations. We have systematically calculated the
binding energy parameter $\lambda$ for both Population I and
Population II stars of masses $1-20 M_{\sun}$, taking into account
the contribution from the internal energy of stellar matter. We
present fitting formulae for $\lambda$ that can be incorporated into
future population synthesis investigations. We also briefly discuss
the possible applications of the results in binary evolutions.
\end{abstract}


\section{Introduction}

One of the key stages in the evolution of close binary stars is the
common  envelope (CE) evolution. When the primary star fills its
Roche-lobe (RL) and the following mass transfer is dynamically
unstable so that the secondary cannot accrete all the transferred
material, the transferred matter will form an envelope embedding
both stars and the binary enters the CE evolution. The secondary
star orbits inside the CE, and experiences  a drag force by the
envelope. This causes the orbital decay and spiral-in of the star
with possible ejection of the envelope \citep[for reviews
see][]{iben93,taa00}. Two main kinds of theories have been developed
to describe the envelope ejection process. One is the $\alpha$
formalism, which considers the conversion of orbital energy to
overcome the envelope binding energy \citep{web84,dek90,dew00}. If
the orbital energy of the secondary is large enough to eject the
envelope, the system will survive and become a compact binary
containing the core of the lobe-filling star and the secondary in a
much smaller orbit; if the orbital energy is not enough, the two
stars will coalesce. The other is the $\gamma$ formalism, which
considers angular momentum transformation during the spiral-in
\citep[see][and reference therein]{nel05}. Despite extensive three
dimensional hydrodynamical simulations of CE evolution \citep[for
recent reviews, see][]{taa00,taa06}, the physics of the CE phase
remains poorly understood, including the efficiency with which the
CE is ejected from the system. Alternative evolutionary paths have
been proposed to avoid the CE evolution, such as those by
\citet{kin99} and \citet{bee07}, who suggested that super-Eddington
accretion may lead to mass ejection from the system instead of
entering the CE stage.

According to \citet{han94, han95} and \citet{dew00}, the total
binding energy of the envelope is described by:
\begin{equation}
\label{equ1} E_{\rm bind} = \int^{M_{\rm donor}}_{M_{\rm core}}
-\frac{GM(r)}{r} {\rm d}m+\alpha_{\rm th}  \int^{M_{\rm
donor}}_{M_{\rm core}} U {\rm d}m,
\end{equation}
where $M_{\rm donor}$ is the mass of the donor star and $M_{\rm
core}$ is its core mass, $-GM(r)/r$ and $U$ are the gravitational
and internal energy of the stellar matter respectively, $G$ is the
gravitational constant, and $\alpha_{\rm th}$ the percentage of the
internal energy contributing to ejection of the envelope, normally
taking the value between 0 and 1.

For convenience the total binding energy of the envelope is usually
expressed as \citep{web84, dek90, dew00}:
\begin{equation}
\label{equ2} E_{\rm bind} = -\frac{G M_{\rm donor} M_{\rm
env}}{\lambda a_{\rm i} r_{\rm L}},
\end{equation}
where $M_{\rm env}$ is the envelope mass, $\lambda$ is the binding
energy  parameter, $r_{\rm L} = R_{\rm L}/a_{\rm i}$ is the ratio of
the RL radius and the orbital separation at the onset of CE, and
$a_{\rm i} r_{\rm L}$ is normally taken as the stellar radius once a
star fills its RL and starts to transfer matter. If we assume that
part of the change of the orbital energy is used to eject the
envelope, as described by \citet{web84}, then
\begin{equation}
E_{\rm bind} = \alpha_{\rm CE} (\frac{GM_{\rm core}M_2}{2a_{\rm f}} -
\frac{GM_{\rm donor}M_2}{2a_{\rm i}})
\end{equation}
where  $\alpha_{\rm CE}$ is the efficiency of converting the orbital
energy to kinetic energy to eject the CE, $M_2$ the mass of the
secondary, and $a_{\rm i}$ and $a_{\rm f}$ refer to the initial and
final orbital separation of the CE phase, respectively.

Combine Eqs.~(2) and (3), the final orbital separation is related to
the pre-CE separation by the following formula,
\begin{equation}
\frac{a_{\rm f}}{a_{\rm i}} = \frac{M_{\rm core} M_2}{M_{\rm donor}}
\frac{1}{M_2+2M_{\rm env}/{\alpha_{\rm CE}\lambda r_{\rm L}}}.
\end{equation}
Equation (4) indicates that the orbital separation after CE is
sensitively dependent on the product of $\alpha_{\rm CE}$ and
$\lambda$, and the convenient way is to treat both $\alpha_{\rm
CE}$ and $\lambda$ as constants of the order of unity. In most of the
population synthesis investigations concerning the evolution of
close binaries in the literature, $\lambda$ was usually taken to be
a constant, normally around $0.5$ throughout the evolution of the
system. However, there have lots of works
\citep[e.g.][]{han94,dew00,dew01, pod03,web07} suggesting that
$\lambda$ is a variable. \citet{dew00} and \citet{dew01} found that
the value of $\lambda$ changes as the star evolves, and reaches far
more than $0.5$ in late evolutionary stages for stars with mass
between $3$ and $6 M_{\odot}$. In recent population synthesis on
post-CE binaries, \citet{dav10} used linear interpolations of the
tabular data in \citet{dew00} to calculate $\lambda$. We notice that
\citet{dew00} and \citet{dew01}'s calculations did not cover Pop. I
stars less massive than $3 M_{\odot}$ and Pop. II stars, and their
results in  tabular form may not be easily used. Obviously more
advanced population synthesis requires a simple description of the
parameter $\lambda$ for better constraints on the initial stellar
population. In this work we are attempting to investigate the
binding energy parameter $\lambda$ by considering the contribution
of  internal energy of stellar matter for both Pop. I and Pop. II
stars, and to find easy-to-use fitting formulae of $\lambda$ for
further research.

This paper is organized as follows. We describe the stellar
evolution code and assumptions adopted in Sect.~\ref{sec2}. In
Sect.~\ref{sec3} we present the calculated results. We summarize our
results and briefly discuss their possible implications in
Sect.~\ref{sec4}.

\section{Model Description}\label{sec2}

We adopt the latest stellar evolution code TWIN\footnote{The code
used here is a variant of the code \textit{ev} described, in its
initial version, by Eggleton (1971, 1972) and Eggleton et al.
(1973), and was developed by the present authors from the version of
Eggleton (2009, unpublished). The current version of \textit{ev} is
obtainable on request from eggleton1@llnl.gov, along with data
tables and a user manual.} developed by \citet{egg71} and modified
later by \citet{egg72}, \citet{egg73} and \citet{pol95}. One of the
major improvements of the code is that, when a low-mass star is
reaching the He flash igniting point, the code will automatically
generate a zero-age horizontal branch star with both the same core
and total mass which enables further evolution.

We take the mixing length parameter $\alpha = l/H_p$ to be 2.0 and
follow the evolution of both Pop. I and Pop. II stars. We assume the
chemical composition of $X=0.7$ and $Z=0.02$ for Pop. I stars and
$X=0.755$ and $Z=0.001$ for Pop. II ones. Once a star evolves off the
main-sequence, we use the layer that contains less than $15\%$ H as
the boundary between the core and the envelope. We then calculate
the envelope binding energy from Eq.~(\ref{equ1}) by setting
$\alpha_{\rm th}$ to be either $0$ or $1$. After that we use
Eq.~(\ref{equ2}) to calculate the corresponding value of $\lambda$,
where we simply take $a_{\rm i} r_{\rm L} = R_{\rm L}$ as the
stellar radius $R$. To include the stellar wind mass loss during the
evolution, we consider the following mechanisms similar to
\citet{hur00}.

1. The modified Nieuwenhuijzen-type wind model \citep{nie90}:
\begin{equation}
\dot{M}_{\rm NJ} = 9.6\times 10^{-15}
(\frac{Z}{Z_{\odot}})^{1/2}R^{0.81}L^{1.24}M^{0.16} M_{\odot}{\rm
yr}^{-1},
\end{equation}
where $M$, $R$, and $L$ are the stellar mass, radius and luminosity
in solar units, respectively. The wind mass loss rate is modified by
the factor $Z^{1/2}$ due to different metallicity \citep{kud87}. The influence 
of metallicity on the strengths of stellar winds arises from the 
fact that the effective number of lines contributing to the line acceleration 
changes with metallicity $Z$ and, hence, causes a metallicity dependence 
of the mass-loss rate. The observations of \citet{puls96}, which in the 
metallicity range between  the Galaxy and the SMC ($Z\sim Z_{\odot}/5$)
 confirms the exponent of $1/2$ for O-stars.

2. The Reimers type wind for giant-branch stars as described by
\citet{kud78} and \citet{ibe83}:
\begin{equation}
\dot{M}_{\rm R} = 4\times 10^{-13} \frac{\eta LR}{M} M_{\odot}{\rm
yr}^{-1},
\end{equation}
with $\eta = 0.5$ to modify the wind loss.

3. The reduced Wolf-Rayet-like mass loss by \citet{ham95} and
\citet{ham98}:
\begin{equation}
\dot{M}_{\rm WR} =  10^{-13} L^{1.5} (1.0 - \mu)  M_{\odot}{\rm
yr}^{-1},
\end{equation}
with
\begin{equation}
\mu = (\frac{M-M_{\rm core}}{M})\min \{5.0,
\max[1.2,(\frac{L}{L_0})^{\kappa}],\}
\end{equation}
where $L_0 = 7.0 \times 10^4$ and $\kappa = -0.5$.

4. AGB stellar wind loss by \citet{vas93}:
\begin{equation}
\log \dot{M}_{\rm VW} = -11.4 + 0.0125 [P_0 -100 \max(M-2.5, 0.0)]
M_{\odot}{\rm yr}^{-1},
\end{equation}
where
 \begin{equation}
\log P_0 = \min (3.3, -2.07-0.9\log M +1.94 \log R),
\end{equation}
and this wind loss is limited to
\begin{equation}
\dot{M}_{\rm VW,max} =  1.36 \times 10^{-9} L M_{\odot} {\rm
yr}^{-1}.
\end{equation}

Then the total mass loss via stellar wind (absolute value) is taken
to be
\begin{equation}
\dot{M} = \max (\dot{M}_{\rm NJ} ,\dot{M}_{\rm R} , \dot{M}_{\rm
WR},\dot{M}_{\rm VW} ).
\end{equation}

When taking into account the contribution of internal energy to
total  binding energy, we include the following terms: the thermal
energy of the stellar matter, the radiation energy, the ionization
of H, H$^{+}$, He, He$^{+}$ and He$^{++}$, the dissociation of
H$_2$, the ionization of C, N, O, Ne, Mg, Si and Fe (these seven
elements are assumed to be fully ionized for all the temperatures).
To descriminate the effect of internal energy, we set $\alpha_{\rm
th}$ to be $0$ and $1$, and calculate the respective binding energy
parameter, namely $\lambda_{\rm g}$ and $\lambda_{\rm b}$,
respectively.

\section{Calculation Results}\label{sec3}
We have made calculations for both Pop. I and Pop. II stars, with
mass  range from 1 to 20 $M_{\odot}$. For each star we follow its
evolution until its age exceeds 15 Gyr or the code crushes for rapid
C burning. Once a star evolves to produce a H-exhausted core we
calculate the binding energy of its envelope and accordingly
$\lambda_{\rm b}$ and $\lambda_{\rm g}$, both with and without the
internal energy of the stellar matter. In Figs.~\ref{fig1} and
\ref{fig2} we show some examples of the evolution of $\lambda_{\rm
b}$ and $\lambda_{\rm g}$ with respect to the stellar radius $R$.

Since a star experiences both expansion and shrinkage throughout its
life, we divide its evolution into three stages after the star
leaves the main-sequence according to the change of the stellar
radius, to make the fitting of $\lambda$ more practicable. Stage 1
begins at the center H exhaustion and ends when the star starts to
shrink (i.e., near center He ignition). Stage 2 follows and ends
when the star starts to expand again\footnote{Generally a star
shrinks after its center He ignition and starts to expand again
later. For convenience of fitting we define the end point of stage 2
(and also the starting point of stage 3) as the minimum of stellar
radius. For example, for Pop I stars which are more massive than
$10M_\odot$, we use $10^{-7}$ center He abundance as the end point
of stage 2. As a less massive star expands again before its center
He exhaustion, we simply assume that stage 2 ends when the
respective HB star starts to expand again.}. Stage 3 begins after that and
continues till the end of the evolution.

Figures \ref{fig1} and \ref{fig2} show that both $\lambda_{\rm b}$
and $\lambda_{\rm g}$ vary during the evolution of stars. For $1
M_{\odot}$ star, both $\lambda_{\rm b}$ and $\lambda_{\rm g}$
decrease with $R$, but their magnitudes are around unity before the
He flash. After that the values of $\lambda$ have a big decrease
along with the stellar radius, but $\lambda_{\rm b} \sim 2
\lambda_{\rm g}$ throughout the evolution. More massive stars often
experience rapid increase in $\lambda$ in stage 3. For stars with
mass from $\sim 3$ to $5M_{\odot}$ (the mass range is related to
metallicity), $\lambda_{\rm b}$ and $\lambda_{\rm g}$ take the
values $\sim 1$ before the star reaches stage 3, during which the
internal energy dominates, and the total binding energy finally becomes
positive. The resulting $\lambda_{\rm b}$ increases rapidly and
eventually turns out to be infinity, while $\lambda_{\rm g}$ remains
to be around 0.5. For stars massive than $6 M_\odot$, the binding
energy decreases in stage 3 but never becomes positive,
correspondingly $\lambda_{\rm b}$ takes smaller value and
$\lambda_{\rm g}$ is around $0.5$. For stars more massive than
$12M_\odot$, $\lambda$ even drops below $0.5$ for most $R$. The
extreme case in our calculation is for $20M_\odot$ stars, in which
both $\lambda_{\rm b}$ and $\lambda_{\rm g}$ evolve to be $\ll 1$ in stage
3.

To provide easy-to-use formulae of $\lambda_{\rm b}$ and
$\lambda_{\rm g}$, we make polynomial fitting with the following
formulation,
\begin{equation}
y = a + b_1x + b_2x^2 +b_3x^3 +b_4x^4 +b_5x^5,
\end{equation}
where $y$ is $\lambda_{\rm b}$ or $\lambda_{\rm g}$, and $x$ is the
stellar radius $R$ in most cases. For low-mass stars, we also use
the fractional mass of the envelope $m_{\rm env} = M_{\rm
env}/M_{\rm donor}$ as the fitting variable as suggested by
\citet{web07}. In limited cases we use $\log \lambda_{\rm b}$
instead of $\lambda_{\rm b}$ as the $y$ variable during stage 3 when
the value of $\lambda_{\rm b}$ is too large, and take $1/\lambda$ as
$y$ and $m_{\rm env}$ as $x$ in several cases (see Tables \ref{tbl1}
and \ref{tbl2} for details). In Tables \ref{tbl1} and \ref{tbl2} we
present the fitting parameters for Pop. I and Pop. II stars,
respectively.

\section{Discussion and conclusions}\label{sec4}

\begin{figure}
\epsscale{.80} \plotone{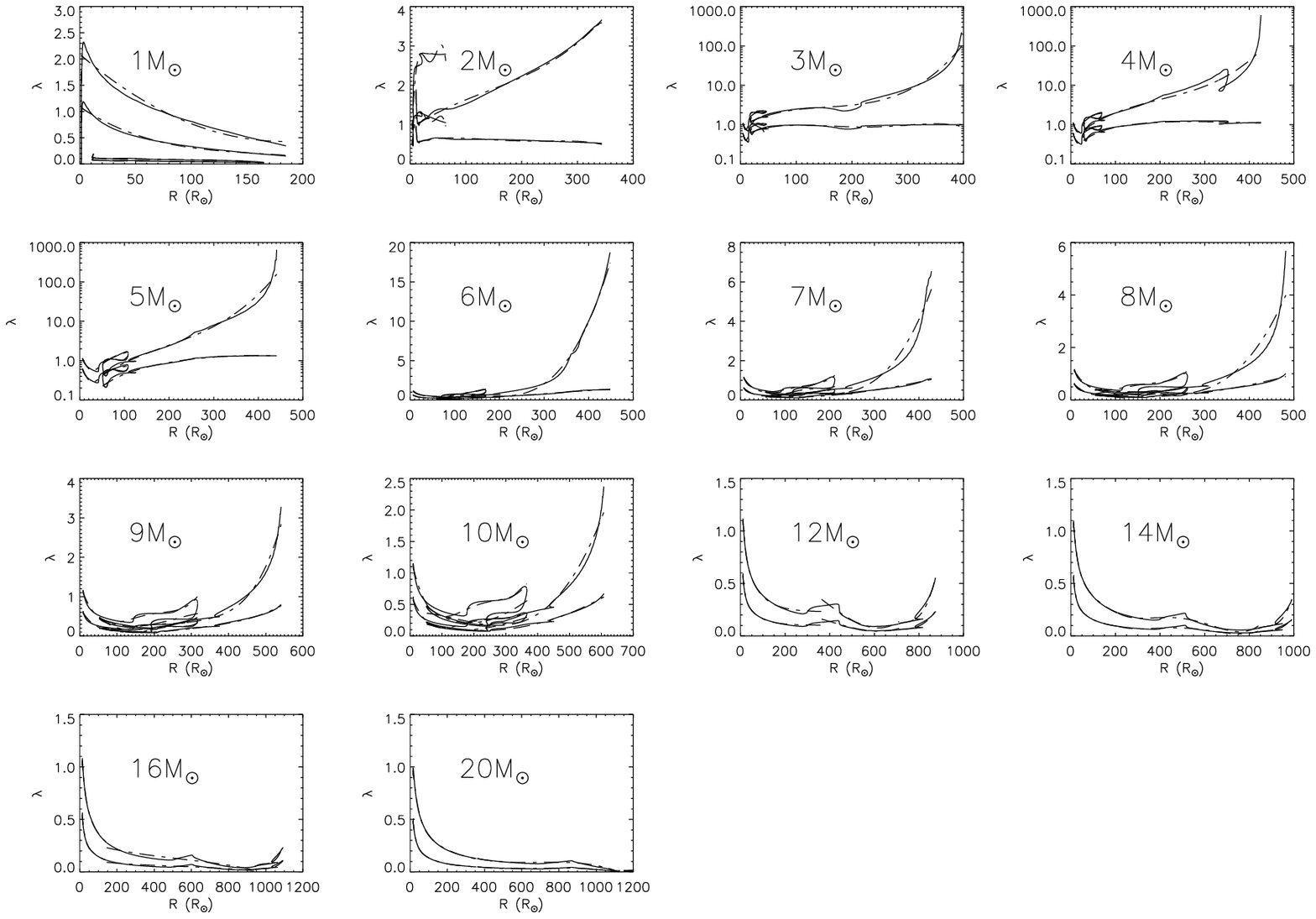} \caption{The evolution of the
binding energy parameter $\lambda$ for Pop. I stars of mass $1
M_{\odot}$ to $20 M_{\odot}$. The upper and lower solid lines are
$\lambda_{\rm b}$ and $\lambda_{\rm g}$, respectively. The
dot-dashed lines are the fitting results. \label{fig1}}
\end{figure}

\begin{figure}
\epsscale{.80} \plotone{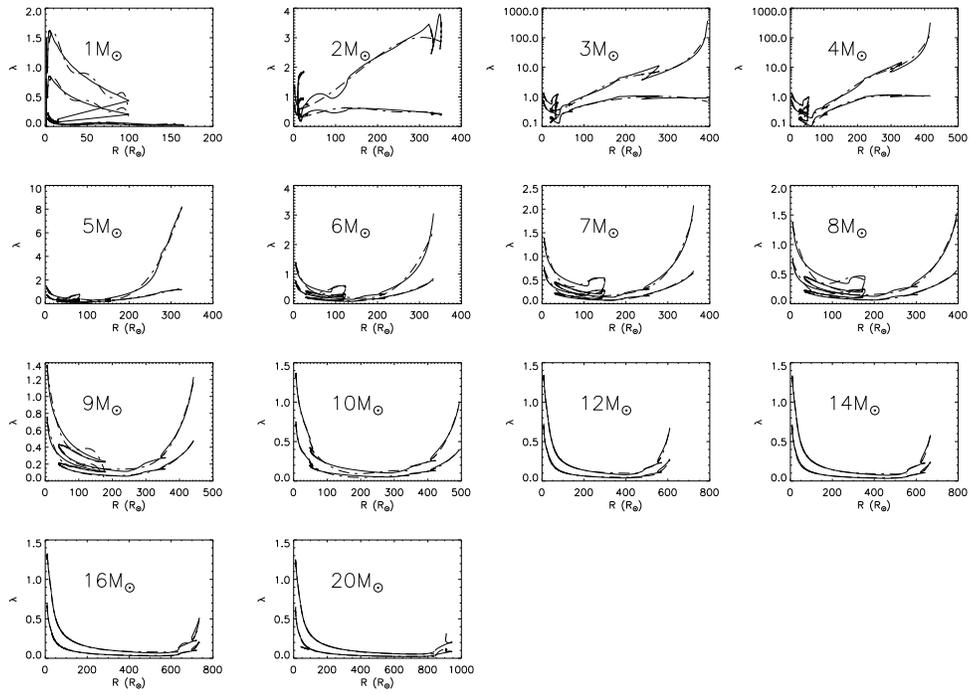} \caption{Same as Fig.~1 but for
Pop. II stars. \label{fig2}}
\end{figure}

Our calculations show that stars with different masses have
different values of $\lambda$, and $\lambda$ is not constant for the
same star in different evolutionary stages. From Figs.~\ref{fig1}
and \ref{fig2}, $\lambda_{\rm b}$ diverges from $0.5$ for most $M$
and $R$, and it is obvious that assuming a constant $\lambda = 0.5$
in population synthesis calculations is far from fact and therefore
lack of reliability. It is also interesting to note that the range
of $\lambda_{\rm g}$ is much narrower than that of $\lambda_{\rm
b}$. Of course the actual value of $\lambda$ should lie between
$\lambda_{\rm g}$ and $\lambda_{\rm b}$ since not all of the internal
energy contributes to the ejection of the envelope \citep{dew00}. As
we have given the fitting formulae for $\lambda$, the results can be
useful for future population synthesis works. However, this
approach should be adopted carefully, especially in the cases when
the binding energies are positive\footnote{Although the
gravitational binding energy remains negative for all the stars
throughout their evolution, the total binding energy, on the other
hand, can turn to be positive for stars with mass ranging from $\sim
3 - 5 M_ {\odot}$ (specifically, $3$ to $5.6 M_{\odot}$ for Pop. I
stars and $2.8$ to $4.5 M_{\odot}$ for Pop. II stars at the end
of their AGB phase when their radii expand to be more than hundreds
of solar radius,
and ionizaiton energy dominates the total binding energy).}, and where
the efficiency of conversion of ionization energy to kinetic energy
may differ from the efficiency of conversion of other forms of
internal energy.
As we can see from Eqs. (\ref{equ1}) and (\ref{equ2}), the
calculated values of $\lambda$ depend on the stellar mass, the core
mass, the envelope mass distribution and the percentage of internal
energy that contributes to envelope ejection. Factors like
metallicity and stellar wind mass loss also affect the stellar
evolution and the resulting values of $\lambda$. Additionally, the
different definitions of core-envelope boundary lead to different
stellar core mass and therefore affect the calculated results.

\begin{figure}
\epsscale{.80} \plotone{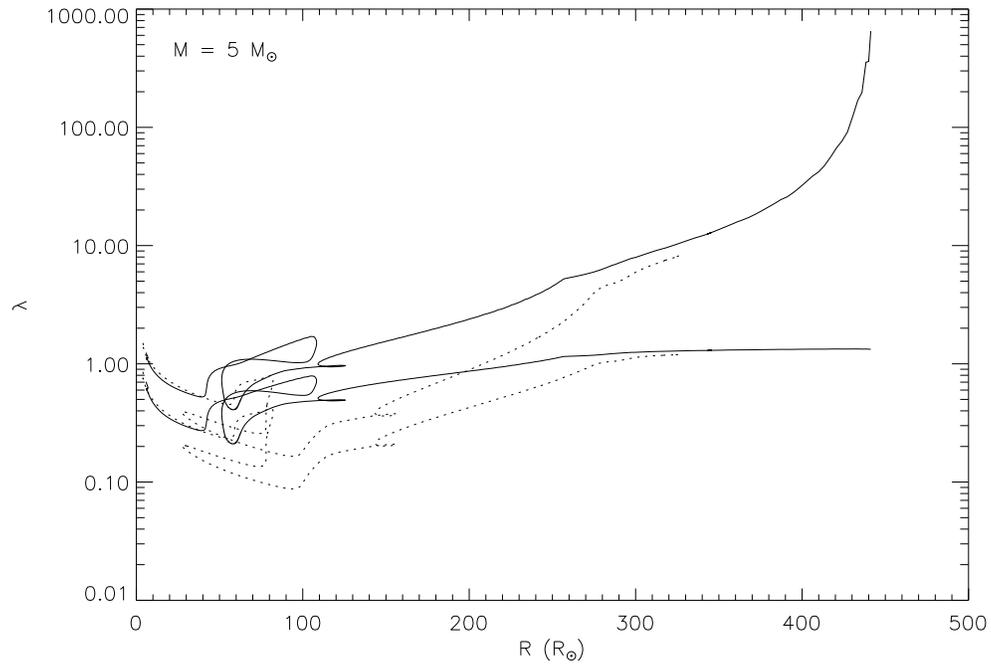} \caption{Comparison of the
$\lambda$ values for $5 M_{\odot}$ Pop. I and Pop. II stars. The
upper and lower solid lines represent $\lambda_{\rm b}$ and
$\lambda_{\rm g}$ for the Pop. I star, respectively. The upper and
lower dotted lines are $\lambda_{\rm b}$ and $\lambda_{\rm g}$ for
the Pop. II star, respectively.\label{fig3}}
\end{figure}

Our results agree reasonably with that of \citet{han94},
\citet{dew00}  and \citet{dew01}, although there exist some
differences. \citet{dew00} proposed that $\lambda$ is almost
independent of the stellar chemical composition. From our
calculations we find that given the same mass, Pop. I and Pop. II
stars have different $\lambda$-values (see Figs.~\ref{fig1} and
\ref{fig2}). For example, the mass range in which the values of
$\lambda_{\rm b}$  can become negative is narrower for Pop. II stars
than for Pop. I ones. In Fig.~\ref{fig3} we compare the evolution of
the two $\lambda$s for a $5 M_\odot$ star with different
metallicity. It is seen that the $\lambda$ evolutions show little
difference at the beginning, but later they diverge from each other,
and the values of $\lambda$ for the Pop. I star always lie above
those of the Pop. II star in late evolutionary stages. The
reason is as follows. The binding energy (absolute value) of the
Pop. I star is either comparable with (in early evolutionary stage) 
or lower than (due to the contribution of ionization energy) that
of the Pop. II star. Since the Pop. I star evolves slower than the
Pop. II one, as a result the Pop. I star has a smaller H-exhausted
core and higher-mass envelope than the Pop. II star when they have
the same radii, thus higher $\lambda$ values. 

We used the 15\% H abundance layer as the stellar core-envelope
boundary to calculate the binding energy and $\lambda$, since the
core mass from such definition increases in most of the time and is
suitable for calculating the binding energy (Eggleton 2009). Note
that \citet{dew00} used the 10\% H abundance layer as the
core-envelope boundary, and in our calculation we found that the two
definitions give similar results. There are other criteria to
define this boundary, such as the maximum energy generation layer or
the density gradient criteria \citep{han94, dew00}. According to
\citet{tau01} and \citet{dew01}, most of these criteria give outer
core boundary and higher core mass. In general, if the total stellar
mass is a constant, a more massive core will decrease the total
binding energy of the envelope and result in a larger
$\lambda$-value \citep{dew00, tau01,dew01}. In other words, the
$\lambda$-values we calculated may be regarded as the lower limit of
the actual ones.
Stars more massive than $10 M_\odot$ lose $10\%$ to $30\%$ of its
total mass during its evolution, which can affect the stellar
structure and the values of $\lambda$ \citep{dew00, dew01, pod03}.
\citet{dew01} and \citet{pod03} have also found that $\lambda$ is
sensitive to the wind mass loss for massive stars, and including
stellar wind leads to higher binding energy (absolute value) and
lower value of $\lambda$.

Our results suggest that $\lambda_{\rm b}$ can take values
significantly larger than $0.5$ during the late-stage evolution of
stars with mass between $\sim 2$ and $10 M_\odot$, which will
significantly increase the efficiency of envelope ejection. If we
consider an evolved giant or AGB star in this mass range with
$\lambda>>0.5$, the envelope binding energy may be much smaller
compared to that with $\lambda = 0.5$. It is then much easier for
its companion to eject the CE, and to produce wider binaries after
the CE phase. It is also worth noticing that both $\lambda_{\rm g}$
and $\lambda_{\rm b}$ are less than $1$ in the late evolutionary
stages of massive stars. For stars massive than $\sim 8 M_{\odot}$,
$\lambda$ even lies below $0.2$. As a result, the actual envelope
ejection efficiency for these stars decreases considerably. Taking a
$20M_\odot$ Pop. I giant star as an example. As seen from
Fig.~\ref{fig1}, both $\lambda_{\rm b}$ and $\lambda_{\rm g}$ are
$\ll 1$, which means that the envelope binding energy become much
higher. The secondary has to convert more orbital energy to eject
the envelope, which probably leads to a much tighter final orbit, or
even a merger. These may have important implications for the
formation and evolution of related binaries that produce compact
stars, which will be discussed in future works in more detail.

\acknowledgments

We thank Prof. P. P. Eggleton for providing the latest version of
the evolutionary code and helpful suggestions. This work was
supported by the Natural Science Foundation of China (under grant
number 10873008) and the National Basic Research Program of China
(973 Program 2009CB824800).


\clearpage

\clearpage

\clearpage

\clearpage

\begin{deluxetable}{ccrrrrrrr}

\tabletypesize{\scriptsize}
\tablecaption{Fitting Results of $\lambda$ for Pop. I Stars
\label{tbl1}} \tablewidth{0pt} \tablehead{ \colhead{Stage} &
\colhead{Mass ($M_\odot$)} & \colhead{$\lambda$}& \colhead{a}
&\colhead{ $b_1$  }& \colhead{$b_2$} & \colhead{$b_3$ }&\colhead{
$b_4$ }&\colhead{ $b_5$}
\\
}
\startdata
1&1\tablenotemark{**}&$\lambda_b$&19.84809&-69.00242&81.24502&-31.5572&0&0\\
& \tablenotemark{**}&$\lambda_g$&26.09283&-72.37166&64.13808&-16.24247&0&0\\
1&2&$\lambda_b$&2.05363&-0.00685&-3.42739E-4&3.93987E-6&-1.18237E-8&0\\
&&$\lambda_g$&1.07658&-0.01041&-4.90553E-5&1.13528E-6&-3.91609E-9&0\\
1&3&$\lambda_b$&2.40831&-0.42459&0.03431&-9.26879E-4&8.24522E-6&0\\
&&$\lambda_g$&1.30705&-0.22924&0.01847&-5.06216E-4&4.57098E-6&0\\
1&4&$\lambda_b$&1.8186&-0.17464&0.00828&-1.31727E-4&7.08329E-7&0\\
&&$\lambda_g$&1.02183&-0.1024&0.00493&-8.16343E-5&4.55426E-7&0\\
1&5&$\lambda_b$&1.52581&-0.08125&0.00219&-2.0527E-5&6.79169E-8&0\\
&&$\lambda_g$&0.85723&-0.04922&0.00137&-1.36163E-5&4.68683E-8&0\\
1&6&$\lambda_b$&1.41601&-0.04965&8.51527E-4&-5.54384E-6&1.32336E-8&0\\
&&$\lambda_g$&0.78428&-0.02959&5.2013E-4&-3.45172E-6&8.17248E-9&0\\
1&7&$\lambda_b$&1.38344&-0.04093&5.78952E-4&-3.19227E-6&6.40902E-9&0\\
&&$\lambda_g$&0.76009&-0.02412&3.47104E-4&-1.92347E-6&3.79609E-9&0\\
1&8&$\lambda_b$&1.35516&-0.03414&4.02065E-4&-1.85931E-6&3.08832E-9&0\\
&&$\lambda_g$&0.73826&-0.01995&2.37842E-4&-1.09803E-6&1.79044E-9&0\\
1&9&$\lambda_b$&1.32549&-0.02845&2.79097E-4&-1.07254E-6&1.46801E-9&0\\
&&$\lambda_g$&0.71571&-0.01657&1.64607E-4&-6.31935E-7&8.52082E-10&0\\
1&10&$\lambda_b$&1.29312&-0.02371&1.93764E-4&-6.19576E-7&7.04227E-10&0\\
&&$\lambda_g$&0.69245&-0.01398&1.17256E-4&-3.81487E-7&4.35818E-10&0\\
1&12&$\lambda_b$&0.23031&0.74955&13.20774&0.63332&68.67896&0\\
&&$\lambda_g$&0.10247&0.31709&60.57985&0.7254&8.90151&0\\
1&14&$\lambda_b$&0.0576&0.59416&158.80613&0.87013&19.04369&0\\
&&$\lambda_g$&0.07178&0.29286&80.4363&0.73414&10.52307&0\\
1&16&$\lambda_b$&-0.2064&0.88945&22.23505&0.79813&287.58284&0\\
&&$\lambda_g$&0.04116&0.74011&11.60388&0.29533&97.68433&0\\
1&20&$\lambda_b$&0.04959&0.85626&29.90918&0.44041&199.20324&0\\
&&$\lambda_g$&0.04101&0.24106&105.06077&0.68656&14.81172&0\\
\\
2&1\tablenotemark{**}&$\lambda_b$&16.71634&-26.38217&-14.18029&0&0&0\\
& \tablenotemark{**}&$\lambda_g$&24.27819&-31.47029&-33.30049&0&0&0\\
2&2&$\lambda_b$&34.41826&-6.65259&0.43823&-0.00953&0&0\\
&&$\lambda_g$&13.66058&-2.48031&0.15275&-0.00303&0&0\\
2&3&$\lambda_b$&-42.98513&7.90134&-0.54646&0.01863&-3.13101E-4&2.07468E-6\\
&&$\lambda_g$&-6.73842&1.06656&-0.05344&0.00116&-9.34446E-6&0\\
2&4&$\lambda_b$&-7.3098&0.56647&-0.01176&7.90112E-5&0&0\\
&&$\lambda_g$&-3.80455&0.29308&-0.00603&4.00471E-5&0&0\\
2&5&$\lambda_b$&-9.93647&0.42831&-0.00544&2.25848E-5&0&0\\
&&$\lambda_g$&-5.33279&0.22728&-0.00285&1.16408E-5&0&0\\
2&6&$\lambda_b$&13.91465&-0.55579&0.00809&-4.94872E-5&1.08899E-7&0\\
&&$\lambda_g$&7.68768&-0.30723&0.00445&-2.70449E-5&5.89712E-8&0\\
2&7&$\lambda_b$&4.12387&-0.12979&0.00153&-7.43227E-6&1.29418E-8&0\\
&&$\lambda_g$&2.18952&-0.06892&8.00936E-4&-3.78092E-6&6.3482E-9&0\\
2&8&$\lambda_b$&-3.89189&0.19378&-0.0032&2.39504E-5&-8.28959E-8&1.07843E-10\\
&&$\lambda_g$&-2.24354&0.10918&-0.00179&1.33244E-5&-4.57829E-8&5.90313E-11\\
2&9&$\lambda_b$&0.86369&-0.00995&4.80837E-5&-6.10454E-8&-2.79504E-12&0\\
&&$\lambda_g$&-0.7299&0.0391&-5.78132E-4&3.7072E-6&-1.07036E-8&1.14833E-11\\
2&10&$\lambda_b$&0.74233&-0.00623&2.04197E-5&-1.30388E-8&0&0\\
&&$\lambda_g$&0.36742&-0.00344&1.27838E-5&-1.0722E-8&0&0\\
2&12\tablenotemark{**}&$\lambda_b$&-7261.84589&43423.48906&-96937.03708&95946.58045&-35560.07607&0\\
& \tablenotemark{**}&$\lambda_g$&-280.84015&1259.61941&-1674.61224&681.20633&0&0\\
2&14\tablenotemark{**}&$\lambda_b$&-3334.04459&20638.88263&-47238.6357&47684.78906&-17966.67435&0\\
& \tablenotemark{**}&$\lambda_g$&-4769.9922&29195.29936&-65789.75807&65234.39892&-24107.00589&0\\
2&16\tablenotemark{**}&$\lambda_b$&-321.34016&1789.56304&-3002.4542&1600.03159&0&0\\
& \tablenotemark{**}&$\lambda_g$&-683.28449&3792.87458&-6376.86784&3419.38525&0&0\\
2&20\tablenotemark{**}&$\lambda_b$&72.31855&482.85039&-2593.27601&3880.69092&-1913.27392&0\\
& \tablenotemark{**}&$\lambda_g$&166.95028&649.50464&-3805.22907&5447.28964&-2487.34665&0\\
\\
3&1\tablenotemark{**}&$\lambda_b$&75.23035&-763.8397&3180.89498&-4665.23406&0&0\\
& \tablenotemark{**}&$\lambda_g$&121.25389&-1169.84175&4686.7113&-6749.8215&0&0\\
3&2&$\lambda_b$&0.88954&0.0098&-3.1411E-5&7.66979E-8&0&0\\
&&$\lambda_g$&0.48271&0.00584&-6.22051E-5&2.41531E-7&-3.1872E-10&0\\
3&3\tablenotemark{*}&$\lambda_b$&-0.04669&0.00764&-4.32726E-5&9.31942E-8&0&0\\
&&$\lambda_g$&0.44889&0.01102&-6.46629E-5&5.66857E-9&7.21818E-10&-1.2201E-12\\
3&4\tablenotemark{*}&$\lambda_b$&-0.37322&0.00943&-3.26033E-5&5.37823E-8&0&0\\
&&$\lambda_g$&0.13153&0.00984&-2.89832E-5&2.63519E-8&0&0\\
3&5\tablenotemark{*}&$\lambda_b$&-0.80011&0.00992&-3.03247E-5&5.26235E-8&0&0\\
&&$\lambda_g$&-0.00456&0.00426&4.71117E-6&-1.72858E-8&0&0\\
3&6&$\lambda_b$&-2.7714&0.06467&-4.01537E-4&7.98466E-7&0&0\\
&&$\lambda_g$&0.23083&-0.00266&2.21788E-5&-2.35696E-8&0&0\\
3&7&$\lambda_b$&-0.63266&0.02054&-1.3646E-4&2.8661E-7&0&0\\
&&$\lambda_g$&0.26294&-0.00253&1.32272E-5&-7.12205E-9&0&0\\
3&8&$\lambda_b$&-0.1288&0.0099&-6.71455E-5&1.33568E-7&0&0\\
&&$\lambda_g$&0.26956&-0.00219&7.97743E-6&-1.53296E-9&0&0\\
3&9&$\lambda_b$&1.19804&-0.01961&1.28222E-4&-3.41278E-7&3.35614E-10&0\\
&&$\lambda_g$&0.40587&-0.0051&2.73866E-5&-5.74476E-8&4.90218E-11&0\\
3&10&$\lambda_b$&0.3707&2.67221E-4&-9.86464E-6&2.26185E-8&0&0\\
&&$\lambda_g$&0.25549&-0.00152&3.35239E-6&2.24224E-10&0&0\\
3&12&$\lambda_b$&-58.03732&0.23633&-3.20535E-4&1.45129E-7&0&0\\
&&$\lambda_g$&-15.11672&0.06331&-8.81542E-5&4.0982E-8&0&0\\
3&14&$\lambda_b$&-106.90553&0.36469&-4.1472E-4&1.57349E-7&0&0\\
&&$\lambda_g$&-39.93089&0.13667&-1.55958E-4&5.94076E-8&0&0\\
3&16&$\lambda_b$&-154.70559&0.46718&-4.70169E-4&1.57773E-7&0&0\\
&&$\lambda_g$&-65.39602&0.19763&-1.99078E-4&6.68766E-8&0&0\\
3&20\tablenotemark{**}&$\lambda_b$&-260484.85724&4.26759E6&-2.33016E7&4.24102E7&0&0\\
&\tablenotemark{**}&$\lambda_g$&-480055.67991&7.87484E6&-4.30546E7&7.84699E7&0&0\\
\\

\enddata

\tablenotetext{*}{Rows with the symbol $*$ uses $\log \lambda_b$
instead of $\lambda_b$ as $y$ variables.}

\tablenotetext{**}{Rows with the symbol $**$ use $m_{\rm env} =
M_{\rm env}/M$  instead of $R$ as $x$ variables and $1/\lambda$ as
$y$ variables.}

\end{deluxetable}

\clearpage

\begin{deluxetable}{ccrrrrrrr}
\tabletypesize{\scriptsize}
\tablecaption{Fitting Results of $\lambda$ for Pop. II
Stars\label{tbl2}} \tablewidth{0pt} \tablehead{ \colhead{Stage} &
\colhead{Mass($M_\odot$)} & \colhead{$\lambda$}& \colhead{a}
&\colhead{ $b_1$  }& \colhead{$b_2$} & \colhead{$b_3$ }&\colhead{
$b_4$ }&\colhead{ $b_5$}
\\
}
\startdata
1&1&$\lambda_b$&1.06031&0.08173&-0.00436&6.7847E-5&-3.3429E-7&0\\
&&$\lambda_g$&0.57192&0.03937&-0.00221&3.49192E-5&-1.73046E-7&0\\
1&2&$\lambda_b$&2.56108&-0.75562&0.1027&-0.00495&8.05436E-5&0\\
&&$\lambda_g$&1.41896&-0.4266&0.05792&-0.00281&4.61E-5&0\\
1&3&$\lambda_b$&1.7814&-0.17138&0.00754&-9.02652E-5&0&0\\
&&$\lambda_g$&0.99218&-0.10082&0.00451&-5.53632E-5&0&0\\
1&4&$\lambda_b$&1.65914&-0.10398&0.0029&-2.24862E-5&0&0\\
&&$\lambda_b$&0.92172&-0.06187&0.00177&-1.42677E-5&0&0\\
1&5&$\lambda_b$&1.58701&-0.06897&0.00129&-6.99399E-6&0&0\\
&&$\lambda_g$&0.87647&-0.04103&7.91444E-4&-4.41644E-6&0&0\\
1&6&$\lambda_b$&1.527&-0.04738&6.1373E-4&-2.36835E-6&0&0\\
&&$\lambda_g$&0.83636&-0.02806&3.73346E-4&-1.47016E-6&0&0\\
1&7&$\lambda_b$&1.49995&-0.03921&4.2327E-4&-1.37646E-6&0&0\\
&&$\lambda_g$&0.81688&-0.02324&2.5804E-4&-8.54696E-7&0&0\\
1&8&$\lambda_b$&1.46826&-0.03184&2.85622E-4&-7.91228E-7&0&0\\
&&$\lambda_g$&0.79396&-0.01903&1.77574E-4&-5.04262E-7&0&0\\
1&9&$\lambda_b$&1.49196&-0.03247&3.08066E-4&-9.53247E-7&0&0\\
&&$\lambda_g$&0.805&-0.02&2.01872E-4&-6.4295E-7&0&0\\
1&10&$\lambda_b$&1.73966&-0.07074&0.00178&-1.73072E-5&0&0\\
&&$\lambda_g$&1.027&-0.05685&0.00164&-1.65206E-5&0&0\\
1&12&$\lambda_b$&1.63634&-0.04646&7.49351E-4&-5.23622E-6&0&0\\
&&$\lambda_g$&1.17934&-0.08481&0.00329&-4.69096E-5&0&0\\
1&14&$\lambda_b$&1.45573&-0.00937&-0.00131&3.07004E-5&0&0\\
&&$\lambda_g$&1.19526&-0.08503&0.00324&-4.58919E-5&0&0\\
1&16&$\lambda_b$&1.33378&0.01274&-0.00234&4.6036E-5&0&0\\
&&$\lambda_g$&1.17731&-0.07834&0.00275&-3.58108E-5&0&0\\
1&20&$\lambda_b$&1.27138&0.00538&-0.0012&1.80776E-5&0&0\\
&&$\lambda_g$&1.07496&-0.05737&0.00153&-1.49005E-5&0&0\\
\\
2&1&$\lambda_b$&0.37294&-0.05825&0.00375&-7.59191E-5&0&0\\
&&$\lambda_g$&0.24816&-0.04102&0.0028&-6.20419E-5&0&0\\
2&2&$\lambda_b$&-103.92538&25.37325&-2.03273&0.0543&0&0\\
&&$\lambda_g$&-56.03478&13.6749&-1.09533&0.02925&0&0\\
2&3&$\lambda_b$&-12.40832&1.59021&-0.06494&8.69587E-4&0&0\\
&&$\lambda_g$&-6.47476&0.8328&-0.03412&4.58399E-4&0&0\\
2&4&$\lambda_b$&-5.89253&0.54296&-0.01527&1.38354E-4&0&0\\
&&$\lambda_g$&-3.21299&0.29583&-0.00833&7.55646E-5&0&0\\
2&5&$\lambda_b$&-0.67176&0.07708&-0.00175&1.1991E-5&0&0\\
&&$\lambda_g$&-0.38561&0.0427&-9.6948E-4&6.64455E-6&0&0\\
2&6&$\lambda_b$&0.30941&0.00965&-2.31975E-4&1.26273E-6&0&0\\
&&$\lambda_g$&0.14576&0.00562&-1.30273E-4&7.06459E-7&0&0\\
2&7&$\lambda_b$&0.44862&0.00234&-9.23152E-5&4.67797E-7&0&0\\
&&$\lambda_g$&0.21873&0.00154&-5.18806E-5&2.60283E-7&0&0\\
2&8&$\lambda_b$&0.50221&-3.19021E-4&-3.81717E-5&1.80726E-7&0&0\\
&&$\lambda_g$&0.24748&-9.9338E-5&-1.99272E-5&9.47504E-8&0&0\\
2&9&$\lambda_b$&0.39342&0.00259&-4.97778E-5&1.69533E-7&0&0\\
&&$\lambda_g$&0.20796&6.62921E-4&-1.84663E-5&6.58983E-8&0&0\\
2&10&$\lambda_b$&0.75746&-0.00852&3.51646E-5&-4.57725E-8&0&0\\
&&$\lambda_g$&0.35355&-0.00388&1.56573E-5&-1.98173E-8&0&0\\
2&12&$\lambda_b$&0.09472&1.59771&19.38009&0.4662&83.08091&0\\
&&$\lambda_g$&0.04621&0.23145&76.88504&0.64758&18.45577&0\\
2&14&$\lambda_b$&0.08576&0.50546&90.52062&1.73793&17.7433&0\\
&&$\lambda_g$&0.04015&0.69617&17.24357&0.23011&86.67988&0\\
2&16&$\lambda_b$&0.0774&1.86228&17.55952&0.52273&99.08578&0\\
&&$\lambda_g$&0.03457&0.73627&17.10365&0.22597&96.30501&0\\
2&20&$\lambda_b$&0.05174&1.52248&25.93257&0.4196&149.18541&0\\
&&$\lambda_g$&0.02187&0.16728&146.24645&0.55973&0&0\\
\\
3&1&$\lambda_b$&0.24012&-0.01907&6.09529E-4&-8.17819E-6&4.83789E-8&-1.04568E-10\\
&&$\lambda_g$&0.15504&-0.01238&3.96633E-4&-5.3329E-6&3.16052E-8&-6.84288E-11\\
3&2&$\lambda_b$&0.5452&0.00212&6.42941E-5&-1.46783E-7&0&0\\
&&$\lambda_g$&0.30594&-9.58858E-4&1.12174E-4&-1.04079E-6&3.4564E-9&-3.91536E-12\\
3&3\tablenotemark{*}&$\lambda_b$&-0.475&-0.00328&1.31101E-4&-6.03669E-7&8.49549E-10&0\\
&&$\lambda_g$&0.05434&0.0039&9.44609E-6&-3.87278E-8&0&0\\
3&4\tablenotemark{*}&$\lambda_b$&-0.2106&-0.01574&2.01107E-4&-6.90334E-7&7.92713E-10&0\\
&&$\lambda_g$&0.36779&-0.00991&1.19411E-4&-3.59574E-7&3.33957E-10&0\\
3&5&$\lambda_b$&-0.12027&0.01981&-2.27908E-4&7.55556E-7&0&0\\
&&$\lambda_g$&0.31252&-0.00527&3.60348E-5&-3.22445E-8&0&0\\
3&6&$\lambda_b$&0.26578&0.00494&-7.02203E-5&2.25289E-7&0&0\\
&&$\lambda_g$&0.26802&-0.00248&6.45229E-6&1.69609E-8&0&0\\
3&7&$\lambda_b$&0.8158&-0.01633&1.46552E-4&-5.75308E-7&8.77711E-10&0\\
&&$\lambda_g$&0.26883&-0.00219&4.12941E-6&1.33138E-8&0&0\\
3&8&$\lambda_b$&0.74924&-0.01233&9.55715E-5&-3.37117E-7&4.67367E-10&0\\
&&$\lambda_g$&0.25249&-0.00161&8.35478E-7&1.25999E-8&0&0\\
3&9&$\lambda_b$&0.73147&-0.01076&7.54308E-5&-2.4114E-7&2.95543E-10&0\\
&&$\lambda_g$&0.31951&-0.00392&2.31815E-5&-6.59418E-8&7.99575E-11&0\\
3&10&$\lambda_b$&-9.26519&0.08064&-2.30952E-4&2.21986E-7&0&0\\
&&$\lambda_g$&0.81491&-0.00161&-8.13352E-6&1.95775E-8&0&0\\
3&12&$\lambda_b$&-51.15252&0.30238&-5.95397E-4&3.91798E-7&0&0\\
&&$\lambda_g$&-13.44&0.08141&-1.641E-4&1.106E-7&0&0\\
3&14&$\lambda_b$&-140&0.7126&0.00121&6.846E-7&0&0\\
&&$\lambda_g$&-44.1964&0.22592&-3.85124E-4&2.19324E-7&0&0\\
3&16&$\lambda_b$&-358.4&1.599&-0.00238&1.178E-6&0&0\\
&&$\lambda_g$&-118.13757&0.52737&-7.8479E-4&3.89585E-7&0&0\\
3&20&$\lambda_b$&-436.00777&1.41375&-0.00153&5.47573E-7&0&0\\
&&$\lambda_g$&-144.53456&0.46579&-4.99197E-4&1.78027E-7&0&0\\

\enddata

\tablenotetext{*}{Rows with the symbol $*$ use $\log \lambda_b$
instead of $\lambda_b$ as $y$ variables.}

\end{deluxetable}

\end{document}